\documentstyle{article}
\def \noi {\noindent}
\def \ssk{\smallskip}

\begin{document}

\title{A principal positioning system for the Earth \footnote{Talk
given at the {\em Journ\'ees Syst\`emes de R\'ef\'erence
Spatio-Temporels} JSR-2002,
 25-28 September, 2002, Bucharest (Romania)}}
\author{Bartolom\'e Coll\\
\small SYRTE/UMR8630-CNRS\\
\small Observatoire de Paris,
61 avenue de l'Observatoire, 75014 Paris, France\\
\small bartolome.coll@obspm.fr, http://coll.cc}
\date{}
\maketitle

\begin{abstract}
 The project SYPOR wishes to use the global
navigation satellite system GALILEO as an autonomous
relativistic positioning system for the Earth. Motivations and a
sketch of the basic concepts underlying the project are
presented. For non geodetic (perturbed) satellites, a
two-dimensional example describes how the dynamics of the
constellation of satellites and that of the users may be deduced
from the knowledge of the dynamics of only one of the satellites
during a partial interval. \end{abstract}

\section{Introduction}

 The current conception of the global
navigation satellite systems (GNSS), like GPS, is based on a
Newtonian model corrected numerically of some "relativistic
effects". The {\em direct} relativistic theory suggests not only
amelioration in accuracy, but also new functions for such GNSS
systems.

\vspace*{1mm}

The project {\em SYPOR} ({\em Sy}st\`eme de {\em Po}sitionnement
{\em R}elativiste) proposes the ideas and instruments needed to
carry out these new possibilities. Particularly it aims to endow
the constellation of satellites of GALILEO of the necessary
elements to constitute, by itself, a {\em primary, autonomous
positioning system} for the Earth and its neighbors. The word
"autonomous" refers here to the capability of the constellation
to provide complete relativistic metric information, i.e. to
describe the {\em kinematics} and the {\em dynamics} both, of
the constellation itself and of the users (possibility of
gravimetry).

\vspace*{1mm}
 For this goal, the project {\em SYPOR} envisages, for the first
time in physics and astronomy, to construct in the neighbors of
the Earth a relativity-compatible physical coordinate system
(relativistic positioning system).

\vspace*{1mm}
 In this paper we sketch the general lines of the
project (Section 2), as well as the underlying physical
concepts, namely that of {\em relativistic positioning systems}
(Section 3) and the way to make such systems {\em autonomous},
which is illustrated in the two-dimensional case (Section 4).

\vspace*{1mm} Some of the  ideas  and  results  on this subject
are the fruit of a long and friendly  collaboration  with
Llu\'is Bel (Univ. Pais Vasco, Bilbao, Spain), Joan Josep
Ferrando and Juan Antonio Morales (Univ. Valencia, Burjassot,
Spain), and Albert Tarantola (Inst. Physique du Globe, Paris,
France).

\section{Sketch of the Project}

For most of the needs of geodesy and positioning, the Earth may
be considered as a {\em Newtonian system}, for which classical
mechanics is enough to explain its essential properties. But a
constellation of satellites around the Earth, endowed with
clocks exchanging their proper times, is a {\em relativistic
system} in its own right (mainly due to Doppler and
gravitational potential "relativistic effects"). Consequently,
the natural conceptual frame to study GNSS is relativity theory.

\vspace*{1mm}

At present, the GNSS involve the Earth and the constellation of
satellites as a sole, coupled system. They start from a
terrestrial, non relativistic coordinate system  (Coll, 2001)
and use the satellites of the constellation as moving beacons to
indicate to the users their position with respect to this
system.

 \vspace*{1mm}

The project SYPOR offers this result in two steps, that have
different levels of conceptual precision and practical accuracy:

\vspace{1mm}
           1. At the first level, SYPOR proposes the concepts and
means to use the sole constellation of satellite-borne clocks as
the most accurate, primary, autonomous, relativistically valid,
positioning system for the neighbors of the Earth. At this
level, any user may know its coordinates with respect to the
satellites, its dynamical state (acceleration, rotation), the
exact internal configuration of the constellation, and their
situation with respect to the ICRS (i.e. all what an user may
hope to obtain from a primary system), and any two users may
know their relative position, distance and relative orientation.

\vspace*{1mm}

     Such a positioning system is a non usual one, with light-like
(rapidly variable) coordinate surfaces, but  any conventional
system may be defined with respect to it. It is to be noted that,
at this level, no synchronization of clocks is at all necessary.

\vspace{1mm}
     2. At the second level, the usual data of the control
segment on the trajectories of the satellites, are "read" as the
data defining the coordinate change to (secondary, non
relativistic) terrestrial coordinates (WGS 84 or ITRF classical
eference systems).

 \vspace*{1mm}

     At this level, any user may know its position (with
terrestrial precision) with respect to the Earth, as with the
current GPS conception.

 \vspace{1mm}

     The possibility appears for a space agency to concentrate its
interest in the first level, the autonomous positioning system,
and delegate to global and local Earth agencies the control of
the terrestrial coordinates.

 \vspace{1mm}
To realize these performances, every satellite must be endowed
with the following kinematic devices:

- a device, on every satellite, allowing to exchange proper
 times with its neighbors  (internal control of the parts of the
system) (Hammesfahr, 1999),

- a device, on  four at least of the satellites, pointing to the
ICRS (International Celestial Reference System) in order to
define virtual local charts "at rest" with respect to the ICRS
(external control of the system as a whole),

- a device, on every satellite, broadcasting over the Earth,
beside its proper time, those of their neighbors (strong
integrity: control by the users segment).

\section{Positioning Systems}

Relativity theory may be used:

* as a wise algorithm to sprinkle Newtonian expressions with terms
corresponding to the "relativistic effects" necessary for the
obtention of the correct numerical values, or

* as the adequate starting frame to rise and to approach the
physical situations with the most recent concepts and
developments on the space-time.

\vspace*{1mm}

The first use is undoubtedly correct in some particular
scenarios, as may be in approximate numerical computations or
for the abstract comparison of the equations of the two
theories. The analysis shows that in fact this first use may be
correct in the situations in which the physical determination of
the coordinates either does not matter, or may be numerically
identified to their Newtonian geometric determinations. But it
is obviously useless to take advantage of the progress and
specific developments of relativity theory in its proper domain,
as it is the case in advanced GNSS.

\vspace*{1mm}
    The basic arena of relativity theory is its space-time.
Relativistic space-time differs from Newtonian space-time in the
following essential point: "the space" and "the present" are not
now "physical objects", but inessential local arbitrary
conventions. The three-dimensional Newtonian space has as much
physical reality as have the Ptolemaic cristal spheres, and
"past", "present" and "futur" are not exhaustive complementary
parts of the space-time. Consequently, "objects" in their usual
sense, like galaxies, stars, planets, mountains  do not exist in
the relativistic space-time. What one can find in it are rather
the "absolute invariants" that they generate, that is to say,
their "histories". It is with the histories of the satellites
that a relativistic theory of GNSS must be constructed.

\vspace*{1mm}
     A {\em locationing system} is a detailed description for
the physical construction of a coordinate system. To construct a
coordinate system is either to construct its coordinate lines,
or to construct its coordinate (hyper-)surfaces. But the choice
of one or the other, and the protocol of their physical
construction, give rise to very different physical properties
(Coll, 2001).

\vspace*{1mm}
     Thus, when the protocol of their physical construction
allows a particular observer (generally situated at the origin)
to attach to every point of his neighbor a set of coordinates,
one has what is currently called a {\em reference system}. But
if this protocol allows every point of a neighbor to know its
proper coordinates, then one has a {\em positionning system}.

\vspace*{1mm} In Newtonian physics, these functions are
exchangeable, and perhaps this is why they are frequently
mistaken. But in relativistic physics these two functions have
very different physical properties:

* they are always {\em incompatible} for a sole coordinate
system. So that the "reference" or "positioning" character of a
locationing system must be previously chosen,

* it is always {\em impossible} to construct a positioning
system starting from a reference system,

* it is always {\em possible} (and very easily) to construct
a reference system starting from a positioning one.

\vspace*{1mm}
    As a consequence, the {\em first} element to be conceived in a
relativistic GNSS must imperatively be its positioning system,
and not its terrestrial reference system (WG 84 or IERS) as it
is currently the case.

\vspace*{1mm}
    Remember (Coll, 2001) that relativistic positioning systems
are {\em generic,} {\em free,} and {\em immediate}, three very
important physical properties that no other locationing system
may simultaneously offer. Thus, for example, Cartesian
coordinates are not generic, the standard synchronization
(two-way signals) usual in the construction of reference systems
does not give rise to immediate systems; harmonic conditions are
generic, but not free, etc.

\vspace*{1mm}
    A general analysis of rigorous mathematical results,
physical possibilities and present technical developments leads
to the important epistemic result that, among all the
relativistic locationing systems,  the set of relativistic
positioning systems exists but constitute a very little class.
And the simplest representative of this class is the one formed
by electromagnetic signals broadcasting the proper time
$\,\tau_i\,$ of four independent clocks $\,S_i\,$ ($\,i = 1, ...
,4\,$).

From now on, we suppose these clocks carried by (not necessarily
geodetic) satellites.

\vspace*{1mm}
 In the space-time, the above wave fronts signals, parameterized by
the proper time of the clocks, draw four families of physical
hypersurfaces moving at the velocity of light, realizing a {\em
covariantly null coordinate system} $\,\{\tau_i\}\,$.

\vspace*{1mm} Coordinate systems of this class are very unusual.
Very different from the current relativistic ones, and still
more from the Newtonian ones, they have been studied by a very
restricted number of specialists in relativity. For an almost
exhaustive bibliography, see (Blagojevi\'c et al., 2001).

\vspace*{1mm} In such covariantly null coordinate systems,
instead of $\,\eta_{\alpha\beta} = diag\{1,-1,-1,-1\}\,,$ the
Minkowski metric of the space-time adopts the complementary
form:

\begin{equation}
 \eta_{\alpha\beta} = \left(  \begin{array}{cccc} \vspace{1mm}
0 & f & g & h \\ \vspace{1mm}
f & 0 & \ell & m \\ \vspace{1mm}
g & \ell & 0 & n\\\vspace{1mm}
h & m & n & 0
\end{array}
\right)
\end{equation}
where $\, f,g,h,\ell,m,n,\,$ are strictly positive functions of
the coordinates $\,\tau_i\,. $

\vspace*{1mm} We see that in such coordinates there is no
time-like asymmetry, the four coordinate surfaces playing
exactly the same role. The nullity of all the diagonal terms in
the above real expression seems to have  erroneously suggested
in the past that such coordinate systems would be "somewhere
degenerate"; this uncorrect intuitive feeling is perhaps the
cause of the absence of studies on them and of their slow
re-discovery by some authors.

\vspace*{1mm} The four proper times $\,\{\tau_i\}\,$ read at a
space-time event by a receptor constitute its (covariantly null)
coordinates with respect to the four satellites.
 But such a system can not be considered as {\em
primary} (with respect to the space-time structure) if we have
not sufficiently information to relate to it any other
coordinate system (Cartesian, harmonic, etc). And for this task,
we need to know the (dynamical) space-time trajectories of the
satellites. In principle, there are many ways to do that, one
simple one being to force satellites to follow prescribed
trajectories, for example geodesics. But the most complete one
is that in which these information is generated and broadcasted
at every instant by the system of satellites itself, {\em
whatever} be their trajectories. When this happens, we call the
primary system {\em autonomous}.

\section{Autonomous Positioning Systems}

 How to make autonomous  such a system of embarked clocks ,
arbitrarily synchronized, broadcasting their proper times? The
answer is very simple: broadcasting, not only their proper time,
but also the proper time of their neighboring satellite's
clocks.

\vspace*{1mm} In other words: let $\,\tau_{ij}\,,$ $\,i \not=
j\,,$ be the proper time of the satellite's clock $\,j\,$
received by the satellite $\,i\,$ at its proper time instant
$\,\tau_i\,.$ Then, the broadcasting of the data
$\,\{\tau_i,\tau_{ij}\}\,$ allow to make autonomous the
covariantly null space-time positioning system $\,\{\tau_i\}\,$.

\vspace*{1mm} Observe that the sixteen data
$\,\{\tau_i,\tau_{ij}\}\,$ received by an observer contains, of
course, the coordinates $\,\{\tau_i\}\,$ ($\,i = 1, ... ,4\,$),
of this observer but also the coordinates
$\,\{\tau_i,\tau_{ij}\}\,$ of {\em every} satellite $\,i\,$ in
the totally covariant null coordinate system that the four
satellites are generating.

\vspace*{1mm} In a four-dimensional Cartesian grid of axis
$\,\tau_i\,$ ($\,i = 1, ... ,4\,$), the data
$\,\{\tau_i,\tau_{ij}\}\,$ received by an user represent at
every instant the proper position of the user as well as the
positions of the satellites. During an arbitrary interval, these
data allow the user to draw in this grid its proper space-time
trajectory as well as the trajectory of the four satellites. Of
course, two different users will found, in their corresponding
grids, that their personal trajectories  are different, but in
the covering domain of proper times, the trajectories of the
four satellites will be necessarily the same.

\vspace*{1mm} Such grids are diffeomorphic to the corresponding
domain of the space-time, but they are not {\em isometric.} The
precise deformation that relates them is nothing but the {\em
metric tensor} $\,g_{ij}(\tau_\ell)\,$ of this domain,  which
gives the dynamical properties of the trajectories (their
inertial and/or gravitational characteristics).

\vspace*{1mm} When the grid trajectories of the satellites are
particularly simple, their space-time dynamical ones are easy to
obtain. For example, in the two-dimensional Minkowski case, if
the grid trajectories of the two satellites are non parallel
straight lines, one can show that necessarily the satellites
follow convergent geodesics, meanwhile if the grid trajectories
are parallel straight lines, the two satellites are necessarily
uniformly accelerated.

\vspace*{1mm} When the grid trajectories are arbitrary, an
important general result may be prove: the knowledge of the
dynamics of {\em one of the satellites} during a {\em relative
acausal interval}, allows to know the dynamics of both of the
satellites as well as that of the user at any later instant.

\vspace*{1mm} In other words: if $\,1\,$ and $\,2\,$ denote the
two clocks, the relative acausal interval of the satellite
$\,1\,$ with respect to the satellite $\,2\,$ at the "instant"
$\,\tau_2\,$ is the interval of the trajectory of the satellite
$\,1\,$ between its proper times $\,\tau'_1 = \tau_{21}\,$ and
$\,\tau''_1\,$ such that $\,\tau_{12} = \tau_2\,.$ Then our
result states that the dynamical knowledge of the satellite
$\,1\,$ during its proper time interval $\,(\tau'_1,\tau''_1)\,$
suffices to know its dynamical trajectory  for any $\,\tau_1
\geq \tau''_1\,$ and of $\,2\,$ and the user for any $\,\tau_2
\geq \tau_{12}\,.$

\vspace*{1mm}
 The relative acausal interval of, say, the
satellite $\,1\,$ with respect to the satellite $\,2\,,$ at the
"instant" $\,\tau_2\,$ is the interval of the trajectory of the
satellite $\,1\,$ between its proper times $\,\tau'_1 =
\tau_{21}\,$ and  $\,\tau''_1\,$ such that $\,\tau_{12} =
\tau_2\,.$ Then, in other words,  our result states that the
dynamical knowledge of the satellite $\,1\,$ during its proper
time interval $\,(\tau'_1,\tau''_1)\,$ suffices to know its
dynamical trajectory  for any $\,\tau_1 \geq \tau''_1\,$ and of
$\,2\,$ and the user for any $\,\tau_2 \geq \tau_{12}\,.$

\vspace*{1mm} Of course, around the Earth more than four
satellites will be convenient. Every four neighboring ones will
constitute a {\em local chart} of the {\em atlas} of covariantly
null coordinate systems enveloping the Earth. From this atlas,
appropriate virtual global conventional coordinate system may be
defined.

\section{References}
\noindent
 {\leftskip=5mm
\parindent=-5mm

\ssk Blagojevi\'c, M., Garecki, J., Hehl, F., Obukhov, Yu.,
2001. Real Null Coframes in General Relativity and GPS type
coordinates, {\em Phys. Rev.} D65 (2002) 044018, e-print:
gr-qc/0110078. Almost exhaustive bibliography.

\vspace*{0.2cm}

Coll, B., 2001. Physical Relativistic Frames, in {\em
Journ\'ees 2001 Syst\`emes de R\'ef\'erence Spatio-Temporels},
Brussels, Belgium, 24-26 September 2001, also http://coll.cc\,.
Some properties of classical and relativistic coordinate systems
are pointed out.

\vspace*{0.2cm}

 Hammesfahr, J. et al., 1999. Intersatellite
Ranging and Autonomous Ephemeris Determination for future
Navigation Systems, Deutsches Zentrum f\"ur Luft- und Raumfahrt
(DLR), document number ISR-DLR-REP-002. Excellent report on the
possibility of intersatellite links (ISL).

\vspace*{0.2cm}

\noi}

\end{document}